\documentclass{bmcart}

\usepackage{amsthm,amsmath}
\RequirePackage[authoryear]{natbib}
\usepackage{url}
\usepackage[utf8]{inputenc} 
\usepackage{graphicx}
\usepackage{mathtools}
\usepackage{xcolor}
\usepackage[normalem]{ulem}

\usepackage{hyperref}

\startlocaldefs
\endlocaldefs

\begin{document}

\begin{frontmatter}

\begin{fmbox}
\dochead{Research}


\title{Platform trials: the impact of common controls on type one error and power}


\author[
  addressref={aff1,aff2},                   
  corref={aff1},                       
  email={QuynhLan.Nguyen@pei.de}   
]{\inits{Q.N.}\fnm{Quynh} \snm{Nguyen}}
\author[
  addressref={aff1},
  email={Katharina.Hees@pei.de}
]{\inits{K.H}\fnm{Katharina} \snm{Hees}}
\author[
  addressref={aff1,aff2},
  email={benjamin.hofner@pei.de}
]{\inits{B.H.}\fnm{Benjamin} \snm{Hofner}}

\address[id=aff1]{
  \orgdiv{Department of Biostatistics},             
  \orgname{Paul-Ehrlich Institut},          
  \city{Langen},                              
  \cny{Germany}                                    
}
\address[id=aff2]{%
  \orgdiv{Department of Medical Informatics, Biometry, and Epidemiology},
  \orgname{Friedrich-Alexander-Universität Erlangen-Nürnberg (FAU)},
  \city{Erlangen},
  \cny{Germany}
}



\end{fmbox}


\begin{abstractbox}

\begin{abstract} 
Platform trials offer a framework to study multiple interventions in a single trial with the opportunity of opening and closing arms. The use of a common control in platform trials can increase efficiency as compared to individual control arms or separate trials per treatment. However, the need for multiplicity adjustment as a consequence of common controls is currently a controversial debate among researchers, pharmaceutical companies, as well as regulators.

We investigate the impact of a common control arm in platform trials on the type one error and power in comparison to what would have been obtained with a platform trial with individual control arms in a simulation study. Furthermore, we evaluate the impact on power in case multiplicity adjustment is required in a platform trial.

In both study designs, the family-wise error rate (FWER) is inflated compared to a standard, two-armed randomized controlled trial when no multiplicity adjustment is applied. In case of a common control, the FWER inflation is smaller. In most circumstances, a platform trial with a common control is still beneficial in terms of sample size and power after multiplicity adjustment, whereas in some cases, the platform trial with a common control loses the efficiency gain. Therefore, we further discuss the need for adjustment in terms of a family definition or hypotheses dependencies.
\end{abstract}




\end{abstractbox}
%

\end{frontmatter}




\section*{Introduction/Background}
In platform trials, multiple treatment arms are evaluated with the possibility to add or drop treatments during the ongoing trial at different time points. A common master protocol for multiple treatment arms can serve as a common logistical, regulatory and methodological framework. In previously conducted systematic literature reviews on master protocols, basket, umbrella and platform trials (\cite{Renfro.2017}, \cite{Park.2019} or \cite{Meyer.2020}), the authors identified an increasing trend of these innovative trial designs, especially in oncology. In addition, platform trials have further gained more attention in the evaluation of COVID-19 treatments in the past years. The flexibility of a platform trial to add and drop treatment arms allows a faster evaluation of new interventions in an ongoing trial. Examples for platform trials are the I-SPY 2 trial \citep{ISPY2.2022}, the STAMPEDE trial \citep{James.2009}, or the RECOVERY trial \citep{RECOVERY.2022} - to name just a few. In addition to potential logistical and operational advantages, the use of a common control is one of the key features in many platform trials. Instead of individual control arms for each treatment, the use of a common control reduces the total sample size in comparison to a trial with individual controls per treatment arm if no multiplicity adjustment is performed. 

However, there is currently no consensus among researchers on the need for adjustment when comparing multiple treatment arms with a common control in the platform trial \citep{Molloy.2022, Wason.2014}. Debates on the need for adjustment in multi-arm trials are not new and have increased with the increase of master protocols and platform trials \citep{Proschan.1995, Freidlin.2008, Stallard.2019, Parker.2020}. 
Advocates for an adjustment argue that the use of a common control leads to dependencies in the test statistics and thus requires adjustment in some trial settings (e.g. in confirmatory settings, or when evaluating different doses of the same treatment) \citep{Hung.2009, Wason.2016}. A random low or high control could lead to more false positive or false negative findings. Especially for regulatory decision making, this dependency could have an impact and should be recognized \citep{Collignon.2020}. A contrary argument to this is that if we examine multiple drugs in separate trials, we do not adjust over these trials and therefore should not adjust in platform trials either. In addition, \cite{Rothman.1990} and \cite{Perneger.1998}, both in favor of no adjustment, add the philosophical question on which tests to include into the family for adjustment regarding the FWER as it seems to be an arbitrary choice. This is further discussed by several authors such as \cite{Hung.2009} and \cite{Frane.2019}. The authors emphasize the importance of the family definition, and recognize that while this definition is to some degree subjective, it is not completely arbitrary. \cite{Proschan.2000} proposed guidance for situations in which adjustment might be necessary based on the design aspects of clinical trials that could lead to multiplicity issues and other aspects such as relatedness of hypotheses or the nature of alternative hypotheses (i.e. if more than one hypotheses need to be significant in order to claim success, the requirement for adjustment differs). Even though the EMA guideline on multiplicity issues \citep{EMA.Multiplicity.2017} mentions the requirement of FWER control as a prerequisite for confirmatory trials, it is also acknowledged that the guideline does not provide an exhaustive discussion of every situation with multiple treatment arms. Specific guidance such as the \textit{Recommendation Paper on the Initiation and Conduct of Complex Clinical Trials} \citep{CTFG.2019}, the guidelines \textit{Interacting with the FDA on Complex Innovative Trial Designs for Drugs and Biological Products}, and \textit{Master Protocols: Efficient Clinical Trial Design Strategies to Expedite Development of Oncology} \citep{FDA.Complex.2020, FDA.Master.2022}, and the \textit{Complex clinical trials - Questions and answers} \citep{EMA.QnA.2022} discuss complex clinical trials such as platform trials. However, no extensive discussion on whether, when and how to adjust for multiplicity in platform trials is included.

Therefore, it is not clear whether testing multiple treatment arms on a platform trial should always require multiplicity adjustment just because the treatments are evaluated under one common master protocol with a common control. Is the mere use of a common control reason enough to require multiplicity adjustment? In fact, the use of a common control in a platform trial even yields a smaller FWER as compared to a platform trial with individual controls per treatment arm. Several authors however, such as \cite{Proschan.1995}, \cite{Howard.2018} and \cite{Bai.2020}, have shown that while the use of a common control reduces the FWER, the $k$-FWER, the probability of at least $k$ false positive findings, is inflated for $k\geq 2$ as compared to individual trials. Due to the use of a common control in platform trials, the corresponding test statistics are correlated and thus affect the FWER and $k$-FWER for $k\geq 2$ in opposite ways. This finding adds an additional aspect to the already complex discussion on whether adjustment is needed or not.

In a simulation study, we evaluate the performance of a platform trial using a common control in terms of different type I error and power concepts in comparison with a platform trial with individual control arms per treatment arm (which in terms of operating characteristics equals a series of separate two-armed randomised controlled trials per treatment under the assumption of independence between the different control arms). We determine the potential error inflation if no multiplicity adjustment is performed as well as the sample size increase in order to restore power after multiplicity adjustment is applied in a platform trial with common controls. Furthermore, we investigate whether an adjustment in platform trials can still be beneficial in terms of sample size and power as compared to running platform trials with individual controls or even individual trials. 

This paper is structured as follows. In the following section we will briefly introduce some notation, as well as several error and power definitions. This is followed by a case study which is the motivation and serves as a basis for the simulation study which is described in the section afterwards. We conclude with a summary and discussion of our findings. In addition, we also elaborate in the discussion part on the need for adjustment due to different sources of multiplicity and hypotheses dependency.

\section*{Methods}

\subsection*{Notation and correlation of test statistics}
\subsubsection*{Individual controls}
We first assume a platform trial with $m$ treatments and individual controls per treatment arm. Let $X_i^j$ denote the outcome of patient $i$, with $i = 1,...,n_j$, treatment $j,\, j = 1,...,m$, and $X_i^{0j}, i=1,...,n_0^j$ the outcome of patient $i$ in the individual control arm for treatment $j$. 
Assuming a normal distributed outcome, i.e., $X_i^j \sim \mathcal{N}(\delta_j,1)$ and $X_i^{0j} \overset{d}{\sim} \mathcal{N}(0,1)$ (for the sake of simplicity, we consider here the variances to be equal to one), the test statistics under $H_0: \delta_j = 0 ~ \forall j$ are defined as 
\begin{align*}
Z_j := \frac{\bar{X^j} - \bar{X^{0j}}}{\sqrt{\frac{1}{n_{j}}+\frac{1}{n_0^j}}} \sim \mathcal{N}(0,1),
\end{align*}
where $\bar{X^j} :=\frac{1}{n_j} \sum_{i = 1}^{n_j} X^j_i$ resp. $\bar{X^j} :=\frac{1}{n_j} \sum_{i = 1}^{n_j} X_i^{0j}$ denotes the arithmetic mean of the patients of arm $j$ reps. of the individual control arm of treatment $j$. 
\subsubsection*{Common control group}
We now assume a platform trial with $m$ treatments and one common control group. Again we assume normal distributed outcomes. Now $X_i^{0} \sim \mathcal{N}(0,1), i=1,...,n_0$ denotes the outcome of patient $i$ in the common control arm and $n_0$ the total sample size of the common control group. For the treatments arms we can stay with the notation above. The platform trial can be flexible in the sense that not all treatments have to start and end recruitment at the same time.  Each treatment is compared to the common control leading again to $m$ hypothesis tests in the platform trial. For the comparison of a treatment arm to the common control, only the concurrently enrolled patients from the control group are used. Therefore, we replace the above notation of $n_0^j$ and $\bar{X^{0j}}$. The latter is now the mean of all control patients that have been concurrently randomized to arm $j$ and $n_0^j:=\# I_j$ now defines the corresponding number of control patients. Furthermore, we define $I_j$ as the subset of control patients $i=1, \ldots, n_0$, which have been concurrently randomized to arm $j$, hence it is $\# I_j=n_0^j$ and $n_0^{j,j'}$ as the number of control patients that have been concurrently randomized either to arm $j$ or to arm $j'$, i.e. $n_0^{j,j'}:=\#\left(I_j \cup I_{j'}\right)$. The test statistic for the comparison of an arm $j=1,\ldots,m$ to the common control group is with the adapted notation analogously defined to the case of individual control groups by
\begin{align*}
Z_j^{*} := \frac{\bar{X^j} - \bar{X^{0j}}}{\sqrt{\frac{1}{n_{j}}+\frac{1}{n_0^j}}} \sim \mathcal{N}(0,1).
\end{align*}
As mentioned above, the use of a common control leads to dependencies among the test statistics. More precisely, the correlation between the two test statistics $Z_j^*$ and $Z_{j'}^*$ for $j,j'=1,\ldots,m$ and $j\neq j'$ is given by 
\begin{align}\label{eq.corr.1}
   \rho_{j,{j^\prime}} &= \frac{1}{\sqrt{\frac{1}{n_j} + \frac{1}{n_{0}^j}}\sqrt{\frac{1}{n_{j^\prime}} + \frac{1}{n_{0}^{j^\prime}}}} \frac{n_0^j + n_0^{j^\prime} - n_0^{j,j'}}{n_0^j n_0^{j^\prime}},
\end{align}
and the joint distribution of $Z_j$ and $Z_{j'}$ is given by a multivariate normal distribution. For the proof of \eqref{eq.corr.1} please have a look at the appendix. If start and end of recruitment is the same for all treatment arms, i.e. $n_0^{j}=n_0^{j'}=n_0$, equation \eqref{eq.corr.1} simplifies to
\begin{align}\label{eq.corr.2}
    \rho_{j,j'} = \frac{1}{\sqrt{\left(\frac{n_0}{n_j} + 1\right)\left(\frac{n_0}{n_{j'}} + 1\right)}},
\end{align}
which matches with which was already shown by \cite{CharlesW.Dunnett.1955}.
If we furthermore assume equal sample sizes for the treatment and control arms (i.e. $n_0=n_j=n_{j'}$), the correlation of the two test statistics results in $0.5$.

\begin{figure}[ht]
	\centering
  \includegraphics[scale=0.35]{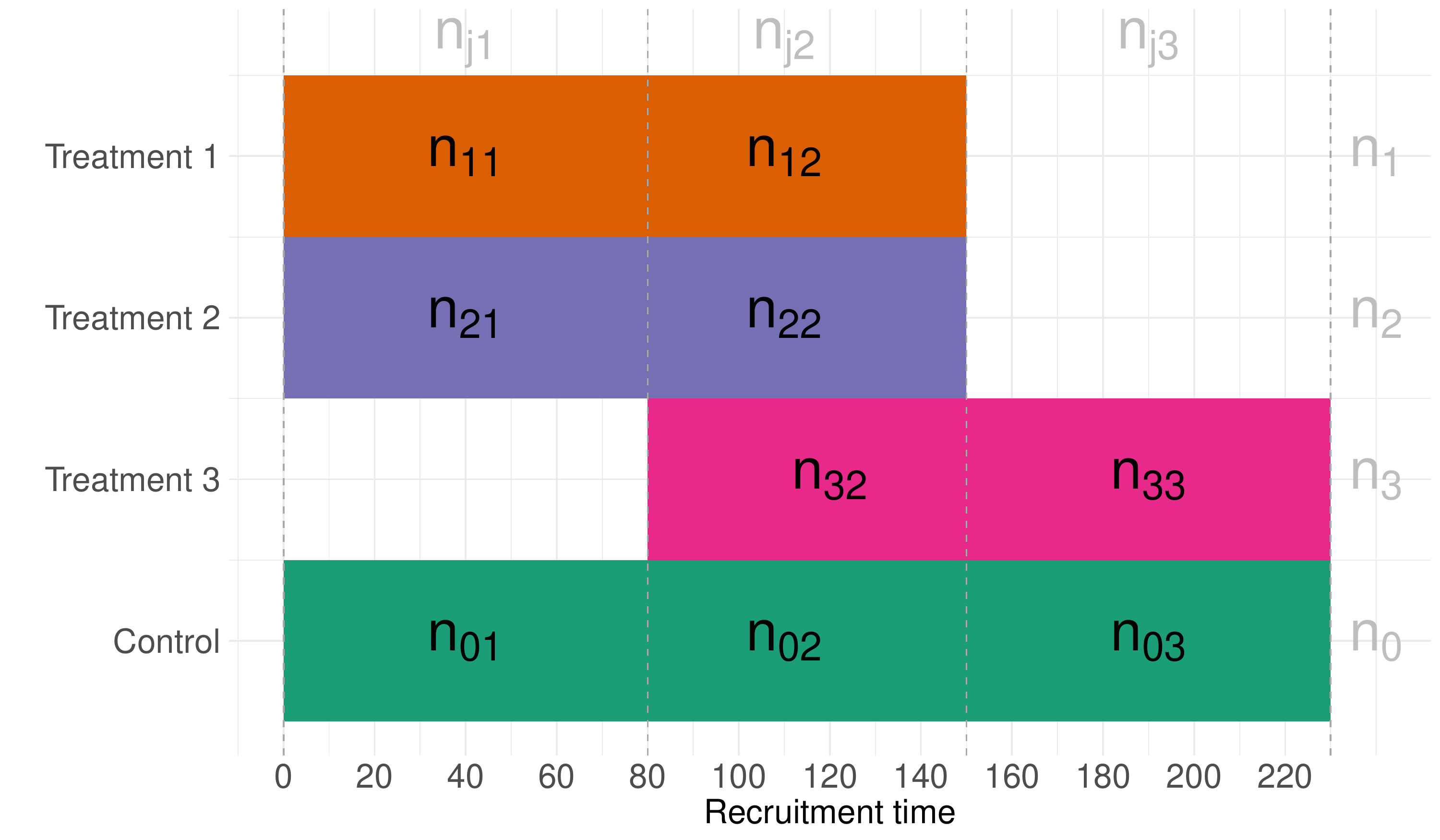}
  \caption{Example of platform trial with three treatment arms of interest and one common control arm}
  	\label{fig:Overlap}
\end{figure}

For illustrative purposes, let's have a look at a concrete example depicted in Figure \ref{fig:Overlap}. Suppose three treatment arms $1$, $2$ and $3$ are compared against a common control. Treatment 3 joins the platform after $n_{01}$ patients are recruited into the control group. Each bar denotes the start and end of recruitment the recruiting period into each arm. The overall sample sizes $n_1$, $n_2$, $n_3$ and $n_0$ of the treatment arms and the control group can be split by the start and end time of the recruitment into recruitment periods such that $n_j = n_{j1} + n_{j2} + n_{j3}$, where $n_{jt}$ denotes the sample size for arm $j$ in period $t$, and $n_{jt} = 0$ if no subjects are enrolled in arm $j$ at period $t$. 

If each treatment arm now is only compared to the concurrently randomized control patients $n_0^j$, the test statistics for treatment 2 compared to the control group and treatment 3 compared to the control group are as follows 
\begin{align*}
Z_2^* = \frac{\bar{X}^2 - \bar{X}^{02}}{\sqrt{\frac{1}{n_{21} + n_{22}} + \frac{1}{n_{01} + n_{02}}}} \sim \mathcal{N}(0,1) \end{align*}
and 
\begin{align*}
Z_3^* = \frac{\bar{X}^3 - \bar{X}^{03}}{\sqrt{\frac{1}{n_{32} + n_{33}} + \frac{1}{n_{02} + n_{03}}}} \sim \mathcal{N}(0,1), \end{align*}
where $\bar{X}^{02}$ denotes the arithmetic mean of the outcomes of the $n_0^2 = n_{01} + n_{02}$ control patients that were concurrently recruited to treatment 2, whereas $\bar{X}^{03}$ denotes the arithmetic mean of the outcomes of the $n_0^3=n_{02} + n_{03}$ control patients that were concurrently recruited to treatment arm 3. It can be seen that the amount of common control patients is now determined by $n_{02}$, leading to the following correlation of $Z_2^*$ and $Z_3^*$: 
\begin{align*}
\rho_{2,3}^* = \frac{1}{\sqrt{\frac{1}{n_{21} + n_{22}} + \frac{1}{n_{01} + n_{02}}}\sqrt{\frac{1}{n_{32} + n_{33}} + \frac{1}{n_{02} + n_{03}}}} \frac{n_{02}}{(n_{01} + n_{02})(n_{02} + n_{03})}.
\end{align*}
If the number of overlapping patients in each recruitment period is the same, i.e. $n_{21} = n_{01}$, $n_{22} = n_{32} = n_{02}$ and $n_{33} = n_{03}$, then the formula for the correlation simplifies to
\begin{align}\label{eq.corr.2}
\rho_{2,3}^* = \frac{1}{2} \frac{n_{02}}{\sqrt{(n_{01} + n_{02})(n_{02} + n_{03})}}.
\end{align}
If start and end of recruitment is the same for all treatment arms with equal sample sizes per arm (i.e. $n_{21} = n_{01} = n_{33} = n_{03}= 0$ and $n_{02}= n_{32} = n_{32}$) then \eqref{eq.corr.2} results in $0.5$, which is consistent with the result in \eqref{eq.corr.1}. \\
Note that the total sample size of a platform trial with a common control is given by $N_\text{CC} = \sum_{j=1}^{m}n_j + n_0$. If instead each treatment is evaluated with individual controls in the platform trial, the total sample size is $N_{IC} = \sum_{j=1}^{m}2\cdot n_j$ when equal allocation is assumed.

\begin{table}[h!]
\caption{Possible outcomes when testing $m$ hypotheses}
\label{tab:error_table}
  \begin{tabular}{lccc}
    \hline
    & $H_0$ true  & $H_0$ false   & Total \\ 
    \cline{2-3}
    Reject $H_0$ & $V$ & $S$ & $R$\\
    Not reject $H_0$ & $U$ & $T$  & $m-R$\\ 
    \cline{2-3}
    Total & $m_0$  & $m-m_0$   & $m$ \\ 
    \hline
  \end{tabular}
\end{table}

\subsection*{Error measures when testing multiple treatment arms}
The classical type I error rate is defined as the probability of rejecting a single null hypothesis $H_0$ given that $H_0$ is true. When testing a family of $m$ hypotheses, we will focus on the following error rates (assuming the notation introduced in Table \ref{tab:error_table}): 
\begin{itemize}
    \item the \textbf{family-wise error rate} (FWER), which is defined as the probability of at least one false positive finding, i.e., $\mathrm{FWER} := P(V \geq 1)$;
    \item the \textbf{$k$-family-wise error rate} ($k$-FWER), also known as the false-multiple error rate, which is defined as the probability of at least $k$- false positive findings, i.e., $k\mathrm{-FWER} := P(V \geq k)$ and hence is a generalization of the FWER;
    \item as well as the \textbf{per-family error rate} (PFER), which is the expected number of false positive findings, i.e., $\mathrm{PFER} := E(V)$.
\end{itemize}
For further details on the error rate definitions, see e.g. \cite{Dmitrienko.2009} or \cite{Hochberg.1987}. 

In order to limit the number of false positive findings, a pre-specified significance level or threshold $\alpha$ can be specified such that the error rate of choice is smaller or equal than $\alpha$. Depending on the situation, control of a specific error rate might be of higher relevance than another. For confirmatory clinical trials for example, the control of the FWER in a strong sense, i.e., regardless of the number of true null hypotheses, is of utmost importance. If all null hypotheses are true, also referred to as the global null (hypothesis), the control of FWER is said to be controlled in a week sense. 

If each single hypothesis is tested at level $\alpha$, it is clear that some error rates can be inflated, e.g., FWER $= 1-(1-\alpha)^m > \alpha$ for $m>2$ under the global null. Methods to control the error rate of choice over the family of $m$ hypotheses include but are not limited to the closed testing procedure, Bonferroni or Dunnett adjustment. Whereby the latter makes use of the correlation between test statistics. 

\subsection*{Power definitions in platform trials}
Besides a type I error, a type II error can be made when testing a hypothesis. The probability of the complementary event, the probability to correctly reject $H_0$ in case the alternative is true and hence the probability to detect a specific treatment effect, is called the power of the trial. The power is in the following denoted by $1-\beta$, where $\beta$ is used for the type II error rate. When testing multiple hypotheses, this power is also referred to as the \textbf{marginal power} for a specific hypothesis test. Similarly to the error rates, the following power concepts can be defined in case of multiple hypotheses tests (using the notation from Table \ref{tab:error_table}):
\begin{itemize}
    \item the \textbf{disjunctive power}, which is defined as the probability to reject at least one false $H_0$, i.e. $P(S \geq 1)$;
    \item as well as the \textbf{conjunctive power}, which is defined as the probability to reject all false $H_0$, i.e. $P(S = m-m_0)$.
\end{itemize}
For further details on the different power concepts, see e.g. \cite{Dmitrienko.2009}.

\section*{Case study}
\subsection*{Motivation and setup}
Our simulation study is motivated by a case study of a fixed four arm platform trial (three treatments and one common control). The phase III trial was proposed as part of an EMA Scientific Advice procedure and details on the background are hence confidential. Each treatment was provided by a different sponsor and was supposed to be compared to the common control and no analysis across the three treatments was planned. The sponsors argued that the evaluation of these treatments under a common master protocol was purely for logistical reason. Therefore, the sponsors proposed no type I error control across the three treatment arms and requested that the results of each treatment should be considered as independent and as if these were three stand-alone clinical trials. As each treatment arm was planned to be tested against a common control and therefore decision-making would not be independent anymore, the control of the type I error (with reference to the multiplicity guideline \citep{EMA.Multiplicity.2017}) over the treatment arms was requested by EMA at that time. 

Both arguments from sponsors and regulators seem reasonable, but are obviously contradicting. In order to quantify the impact of the common control on the type I error rate and on the overall decision making, we performed a small simulation study.

Following the proposed study design, we assumed $n=150$ patients per arm, leading to a total sample size of a platform trial with a common control of $N_{CC} = 4 \cdot 150 = 600$ patients. If the treatments are explored with individual controls in the platform trial (or analogously even in separate trials) a total sample size of $N_{IC} = 3 \cdot 2 \cdot 150 = 900$ patients is needed to achieve the same marginal power. Under the global null, we assumed independently, normally distributed outcomes with mean 0 and variance 1 for each treatment and control arm. Each hypothesis test was performed at a two-sided significance level of $\alpha = 0.05$. For the platform trial with a common control, the error rates were evaluated for the following three scenarios: without the application of any multiplicity adjustment, with Bonferroni adjustment and with Dunnett adjustment. No multiplicity adjustment was considered in the case of individual controls.

\subsection*{Extension to a flexible design}
In a flexible platform trial, one or more treatment arms can join the platform at later time points. At the time of joining, other treatment arms and the control group have potentially already started recruitment.
As an example, we modified our case study such that the third treatment arm started recruitment at a later point in time. Table \ref{tab:Shift_Recruitment1} depicts a situation where treatment 3 joins the platform after 80 patients per arm were already recruited. The number of common control patients for treatment 3 with the other treatments is 70 patients, whereas treatment 1 and treatment 2 share 150 control patients. Note that now the total sample size in the platform trial with the common control was increased by the additional control group patients that were recruited concurrently to the third treatment after the first two treatments stopped recruitment, i.e. $N_{CC}^*=N_{CC}+80=600+80=680$. This shows that in a scenario where one expects equal allocation to each single active arm and the control arm for any given time, flexible platform trials with a common control lead to an increased sample size. We conducted also for this scenario a simulation similar to the one above and compared the results with a flexible platform trial with individual controls, in which each treatment and control group consisted of 150 patients. 
\begin{table}[h!]
\caption{Example of a flexible platform trial with three treatments and a common control. Treatment 3 started recruitment after 80 patients per arm were already recruited to the platform.}
\label{tab:Shift_Recruitment1}
  \begin{tabular}{ccccc}
    \hline
    & $n_{j1}$ & $n_{j2}$ & $n_{j3}$ &  \\ 
    \cline{2-4}
    Treatment 1    & 80 & 70 & 0  & 150\\
    Treatment 2    & 80 & 70 & 0  & 150\\
    Treatment 3    & 0  & 70 & 80 & 150 \\
    Common Control & 80 & 70 & 80 & 230 \\ 
    \cline{2-4}
                  & 240 & 280 & 160 & 680 \\
    \hline
  \end{tabular}
\end{table}

\subsection*{Results}
The simulation results of our motivational case study are provided in Table \ref{tab:scen1_error}. In both study designs with and without a common control, the platform-wise FWER is clearly inflated (i.e. $> 0.05$) if no multiplicity adjustment is applied. While the FWER is smaller with a common control ($0.1251$ vs.\ $0.1403$ without common control), the k-FWER ($k\geq 2$) is increased in comparison to individual controls without adjustment. This can be explained by the use of a common control in the platform trial, which leads to dependent test statistics: If a random low in the common control results in one false positive finding, the likelihood of further false positive findings is increased and is higher as compared to a platform trial with individual controls. The random low equally affects all arms and hence the probability of at least $k$ false positive findings increases. On the other hand, a randomly high common control reduces the probability of at least one false positive finding due to the shared use of this control in each test. The PFER remains the same in the two study designs which is basically due to the fact that the expected value of the sum of several random variables is equal to the sum of their expectations. Even though a platform trial with a common control has a higher chance of making \textit{multiple} false positive findings, \textit{on average} the number of false positive findings is the same as compared to a platform trial with individual controls \citep{Proschan.1995,Howard.2018,Bai.2020}. 

As expected, the FWER is controlled at $0.05$ when Bonferroni or Dunnett adjustment is used. The latter incorporates the correlation of test statistics and is thus less strict than Bonferroni and better exhausts the significance level. It hence leads to an increased power as we will also discuss in the next section.    
\begin{table}[ht]
\caption{Error rates in a four arm platform trial with a common control versus a platform trial with individual controls under the global null. Results without and with multiplicity control are provided for the platform trial with common control 
\label{tab:scen1_error}} 
\begin{center}
\begin{tabular}{lcccccc}
\hline
\multicolumn{1}{c}{\bfseries }&\multicolumn{1}{c}{\bfseries }&\multicolumn{3}{c}{\bfseries Common control}&\multicolumn{1}{c}{\bfseries }&\multicolumn{1}{c}{\bfseries Individual controls}\tabularnewline
\cline{3-5} \cline{7-7}
\multicolumn{1}{l}{} &\multicolumn{1}{c}{}&\multicolumn{1}{c}{Unadjusted}&\multicolumn{1}{c}{Bonferroni}&\multicolumn{1}{c}{Dunnett}&\multicolumn{1}{c}{}&\multicolumn{1}{c}{Unadjusted}\tabularnewline
\cline{2-7}
FWER, P(V$\geq$1)&& 0.1247 & 0.0436 & 0.0489 && 0.1400\tabularnewline
2-FWER, P(V$\geq$2)&& 0.0207 & 0.0046 & 0.0056 && 0.0073\tabularnewline
3-FWER, P(V$\geq$3)&& 0.0030 &0.0005 & 0.0007 &&0.0001\tabularnewline
PFER, E(V)&& 0.1485 & 0.0486 & 0.0552 && 0.1475\tabularnewline
\hline
\end{tabular}\end{center}
\end{table}

We further considered a flexible platform trial and extended our case study. Table \ref{tab:scen1_flex_error} shows the error rates for the flexible platform trial when 80 subjects per arm have already been recruited to the platform when the sponsor of the third arm joins. In both settings with a common control or with individual controls an inflation in the FWER and $k$-FWER is observed if no multiplicity adjustment is included. Since not all treatments started at the same time, the number of common control patients reduces and thus, the FWER in the flexible platform trial with a common control is higher as compared to the fixed platform with common control. The opposite is observed for the $k$-FWER, $k \geq 2$ which is smaller in a flexible platform trial with a common control. With decreasing number of common controls, the correlation between the test statistics decreases and thus, the error rates in the platform trial with common control becomes closer to the error rates obtained with individual controls. Furthermore, we see that the error rates were basically the same for the fixed and flexible platform trial with individual controls (neglecting small differences due to simulation error).    
\begin{table}[ht]
\caption{Error rates in a four arm flexible platform trial with a common control versus a platform trial with individual controls under the global null when 80 patients per arm are already on the trial. Results without and with multiplicity control are provided for the platform trial with common control \label{tab:scen1_flex_error}} 
\begin{center}
\begin{tabular}{lcccccc}
\hline
\multicolumn{1}{c}{\bfseries }&\multicolumn{1}{c}{\bfseries }&\multicolumn{3}{c}{\bfseries Common control}&\multicolumn{1}{c}{\bfseries }&\multicolumn{1}{c}{\bfseries Individual controls}\tabularnewline
\cline{3-5} \cline{7-7}
\multicolumn{1}{l}{} &\multicolumn{1}{c}{}&\multicolumn{1}{c}{Unadjusted}&\multicolumn{1}{c}{Bonferroni}&\multicolumn{1}{c}{Dunnett}&\multicolumn{1}{c}{}&\multicolumn{1}{c}{Unadjusted}\tabularnewline
\cline{2-7}
FWER, P(V$\geq$1) &&$0.1360$&$0.0463$&$0.0495$&&$0.1411$\tabularnewline
2-FWER, P(V$\geq$2) &&$0.0148$&$0.0029$&$0.0033$&&$0.0073$\tabularnewline
3-FWER, P(V$\geq$3) &&$0.0010$&$0.0002$&$0.0002$&&$0.0001$\tabularnewline
PFER, E(V)&&$0.1518$&$0.0493$&$0.0530$&&$0.1486$\tabularnewline
\hline
\end{tabular}\end{center}
\end{table}

\section*{Simulation study}

Starting from the case study, we expanded the simulation to further evaluate the impact of a common control on the type I error rate and power as well as the sample size in comparison to platform trials without common controls. In order to evaluate the impact on different operational characteristics, in each simulation we varied one or more parameters of interest while all other parameters remained fix. For each scenario $M=50.000$ simulations were performed.


For the platform trial with a common control, we explored two multiplicity adjustments (Dunnett and Bonferroni) in addition to no adjustment. In the platform trials with individual controls no adjustment over treatment arms was performed, since no data is shared. Hence, these settings can be considered similarly to performing individual clinical trials per treatment arm with the mere difference of a common randomization framework. 

\subsection*{Fixed platform trial}
For the fixed platform trial, where all treatment arms start and end recruitment at the same time, we mainly focused on the change in error and power rates with increasing number of treatment arms. For each scenario, we simulated the outcome as 
\begin{align}
    X_i^j \; \overset{\mathrm{iid}}{\sim} \; \mathcal{N}(\delta_j,1) \quad i=1, \ldots, n_j, \; j = 1, \ldots, m, \label{eq:sim_treatment}\\
     X_i^{0j} \; \overset{\mathrm{iid}}{\sim} \; \mathcal{N}(0,1) \quad i =, \ldots, n_0^j, \; j = 1, \ldots, m \text{, and} \label{eq:sim_ind_contr}\\
    X_i^0 \; \overset{\mathrm{iid}}{\sim} \; \mathcal{N}(0,1) \quad i=1, \ldots, n_0, \label{eq:sim_com_contr}
\end{align}
for the treatment arms, individual control arms and common control arm, respectively. By changing the effect sizes $\delta_j$ in the treatment arms, we were able to simulate the different cases below. Under the global null, we assumed $\delta_j = 0 ~ \forall j$, assuming each treatment to be ineffective.

\subsubsection*{Null case}
In order to evaluate the impact of a common control on the relevant error rates, we simulated platform trials under the global null with a common control and with individual controls.

We varied the number of treatment arms $m$ between 2 and 10. For both settings, we assumed $n_j = n_0^j = n_0 = n = 150$ per arm, for $j = 1, \ldots, m$, leading to a total sample size of $N_{CC} = n (m + 1)$ in the platform trial with a common control and $N_{IC} = n (2 \cdot m)$ with individual controls. 


\subsubsection*{Single effective case} In order to assess the impact of the study design (individual or common control), the number of arms, and multiplicity correction on the required sample size, we simulated one effective treatment while the control groups and other treatment arms remained ineffective. Without loss of generality, we simulated $\delta_1 = 0.38$, assuming treatment 1 to be effective and $\delta_j = 0, j = 2, \ldots, m$ for the remaining treatment arms. The effect size was chosen such that it mimics the case study as in a two-arm trial, $n=150$ patients per arm are needed to detect a treatment effect of $0.38$ with a power of about 90\% and a two-sided significance level of 5\%.  
We varied the number of treatment arms $m = 2,...,10$. The sample sizes per treatment and control arm $n = n_j = n_0^j = n_0$, for $j = 1, \ldots, m$, were chosen such that a marginal power of 90\% was obtained for the effective treatment arm under each scenario. 

In practice, an arbitrarily large increase in the sample size in clinical trials may not always be feasible. Reasons can be for example constrains in available patients, or budget. Hence, we further explored the marginal power, when the total sample size is fixed at $N_{CC} = N_{IC} = 600$. With varying number of treatment arms, this leads to varying sample sizes per treatment arm. In a platform trial with common control $n_j = n_0 = 600/(m+1)$ and in a platform trial with individual controls $n_j = n_0^j = 600/(2\cdot m)$. This means that the sample sizes per arm varied between $200$ and $54$ for trials with a common control and between $150$ and $30$ for trials with individual controls. 

\subsubsection*{All effective case} In the scenario with fixed sample size ($N_{CC}=N_{IC}=600$), we additionally evaluated the disjunctive and conjunctive power with increasing number of treatment arms $m=2,...,10$. For these simulations, we assumed that all treatments are effective and therefore simulated $\delta_j = 0.38 ~ \forall j$. 

\subsection*{Flexible platform trial}
Here, we mainly focus on how the time of joining a platform trial with a common control affects the type 1 error and power rates and therefore restricted our simulations to $\leq 3$ treatment arms for simplicity. We contrast these results with a flexible platform with individual control arms. For each scenario, we simulated the outcome analogously to Equations~\eqref{eq:sim_treatment} to~\eqref{eq:sim_com_contr} with adapted sample sizes per arm depending on the time of joining. Of note, in a trial with individual control arms it is irrelevant whether the platform is a fixed or flexible platform design as subtrials consisting of a treatment arm and its control are not affected by the opening of new arms, at least in an ideal world. In practice, the new treatment(s) might change the willingness of subjects to join the trial and hence might impact the overall patient population. This is not part of this research, however.

\subsubsection*{Null case} In order to evaluate the effect of a common control and the time of joining a platform on the error rates, we used the example above (Table~\ref{tab:Shift_Recruitment1}) with $m=3$ treatments and assumed that the third treatment arm joins the trial later. Under the global null we assumed for each treatment-control comparison 150 vs. 150 patients, i.e. $n_1 = n_2 = n_3 = 150$ for all treatments and $n_{01} + n_{02} = n_{02} + n_{03} = 150$ for the common control group (see Figure~\ref{fig:Overlap} for definition of sample sizes). We varied the number of patients $n_{j1}$ for the first two treatment arms and the common control group from 0 to 150 patients. This represents the time when treatment 3 joins the platform trial, with the extreme cases of arm 3 running in parallel to arms 1 and 2 ($n_{j1} = 0$) and arm 3 running after arms 1 and 2 completed recruitment, i.e., independently ($n_{j1} = 150$). The results were compared to the design with individual controls, which was simulated as a flexible platform trial with $n_j = n_0^j = 150,\; j = 1, 2, 3$.

\subsubsection*{Single effective case} We followed the same approach as for the single effective case in the fixed design and assumed only one effective treatment with $\delta_3 = 0.38$, while the other treatments were ineffective ($\delta_j = 0, j = 1, 2$). We further defined that the effective treatment arm joins the platform with a common control later. Similarly to above, we varied the number of patients $n_{j1}$ for the first two treatment arms and the common control group from 0 to 150 patients. The required sample size per arm was derived such that a marginal power of 90\% was obtained for the effective arm. We compared the total sample size to the design with individual controls. In the latter, we simulated a flexible platform trial with $n_j = n_0^j = n = 150,\; j = 1, \ldots, m$ such that a marginal power of 90\% was obtained for the effective treatment arm. 

As before, we also considered a scenario with restricted resources. It is clear that if we fix a total sample size $N_{CC}$ over the flexible platform similarly to above, the sponsor that joins the platform later loses on power the later the sponsor joins the platform with the common control. The later the third sponsor joins the trial the greater the number of patients in the first period $n_{j1}, j=0,1,2$ (see Table \ref{tab:Shift_Recruitment1}) which the sponsor cannot use. Since the total sample size is fixed, this only reduces the comparison-wise sample size for the sponsor's treatment against the control. Therefore, we considered a scenario in which the third sponsor has fixed resources and can either spend them in a platform trial with individual controls (or if we ignore the increased overhead costs in a separate trial) or an ongoing platform with a common control.  
We assumed that the sponsor of the newly added treatment arm has a limit of 300 patients. In a trial with individual controls and equal allocation, this results in 150 patients per group. In a flexible platform trial with a common control, the sponsor can take advantage of the common control. Table~\ref{tab:Shift_Recruitment} shows an example, where the third sponsor joins the platform with a common control, when already 90 patients per arm have been recruited to the other treatment arms. We assumed that the third sponsor equally contributes to the $n_{02} = 60$ common control patients who are enrolled in period 2, i.e. is in charge of $60/3$ patients. Hence, sponsor 3 can allocate $300-60/3$ patients to the third treatment in period 2 and 3 ($n_{32}$ and $n_{33}$) as well as the control arm in period 3 ($n_{03}$) such that equal number of patients will be obtained for the treatment-control comparison. We investigate the impact of this budget-driven sample size choice on the marginal power of sponsor 3. For our simulation we varied the number of patients already on the trial $n_{j1}$ for the first two treatments and the common control from 0 to 150 when the third treatment joins the platform trial with a common control. For the design with individual controls, we simulated a flexible platform trial with 150 patients per arm. 
\begin{table}[h!]
\caption{Example of flexible platform trial with three treatments and a common control. Treatment 3 starts recruitment after 90 patients have already been recruited per arm to the platform. The asterix (*) denotes the 300 subjects enrolled by sponsor 3.}
\label{tab:Shift_Recruitment}
  \begin{tabular}{ccccc}
    \hline
    & $n_{j1}$ & $n_{j2}$ & $n_{j3}$ & \\ 
    \cline{2-4}
    Treatment 1 & 90 & 60 & 0 & 150 \\
    Treatment 2 & 90 & 60  & 0 & 150 \\
    Treatment 3 & 0 & $60^{*}$ & $110^{*}$ & 170  \\
    Common Control & 90  & $40+20^{*}$ & $110^{*}$ & 260 \\ 
    \cline{2-4}
     & 270 & 240 & 220 & 730 \\ \hline
  \end{tabular}
\end{table}

\subsubsection*{All effective case} We evaluated the two power concepts for $m=3$ treatments and a comparison-wise sample size of 300 patients in the budget-driven scenario. We assumed all treatments to be effective $\delta_j= 0.38,\; j = 1,2,3$ and varied $n_{j1}$ from 0 to 150 as mentioned above. The flexible platform trial with individual control groups was simulated with 150 vs. 150 patients. 


\section*{Results}
\subsection*{Fixed platform trial}
\subsubsection*{Null case: Comparison of error rates}
\begin{figure}[ht]
	\centering
  \includegraphics[scale=0.35]{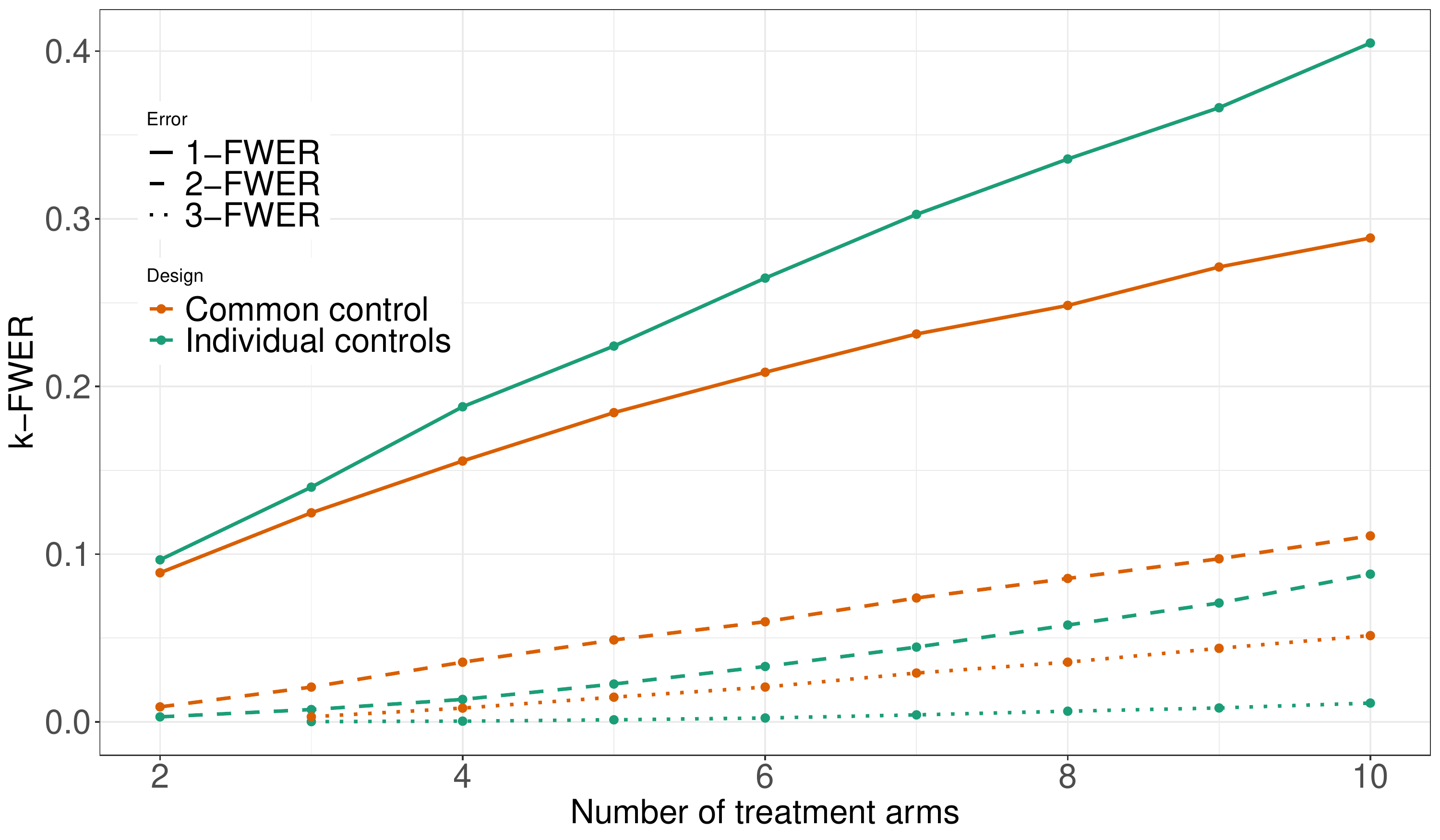}
  \caption{$k$-FWER ($k=1,2,3$) in a platform trial with $j=2,...,10$ treatment arms versus a common control (green) and individual controls (orange) without multiplicity adjustment.}
  	\label{fig:kFWER}
\end{figure}

In Figure \ref{fig:kFWER}, the FWER inflation as well as the $2$-FWER and $3$-FWER increase for platform trials with and without common controls under the global null are shown for increasing number of treatment arms. While the difference in FWER inflation between the two designs is small in our case study with $j = 3$ treatment arms, the difference increases with increasing number of treatment arms. The opposite is observed for the $2$-FWER and $3$-FWER. Numerically the $2$-FWER is much smaller than the FWER but exceeds 0.05 with $j=5$ in the platform trial with a common control and $j=8$ with individual controls. When Bonferroni or Dunnett adjustment is performed, the FWER is controlled and therefore also the $k$-FWER is below 0.05 and ss discussed in the case study, the PFER is identical for the two designs (results not shown here).     

\subsubsection*{Single effective case: Comparison of marginal power}
\begin{figure}[ht]
	\centering
  \includegraphics[scale=0.23]{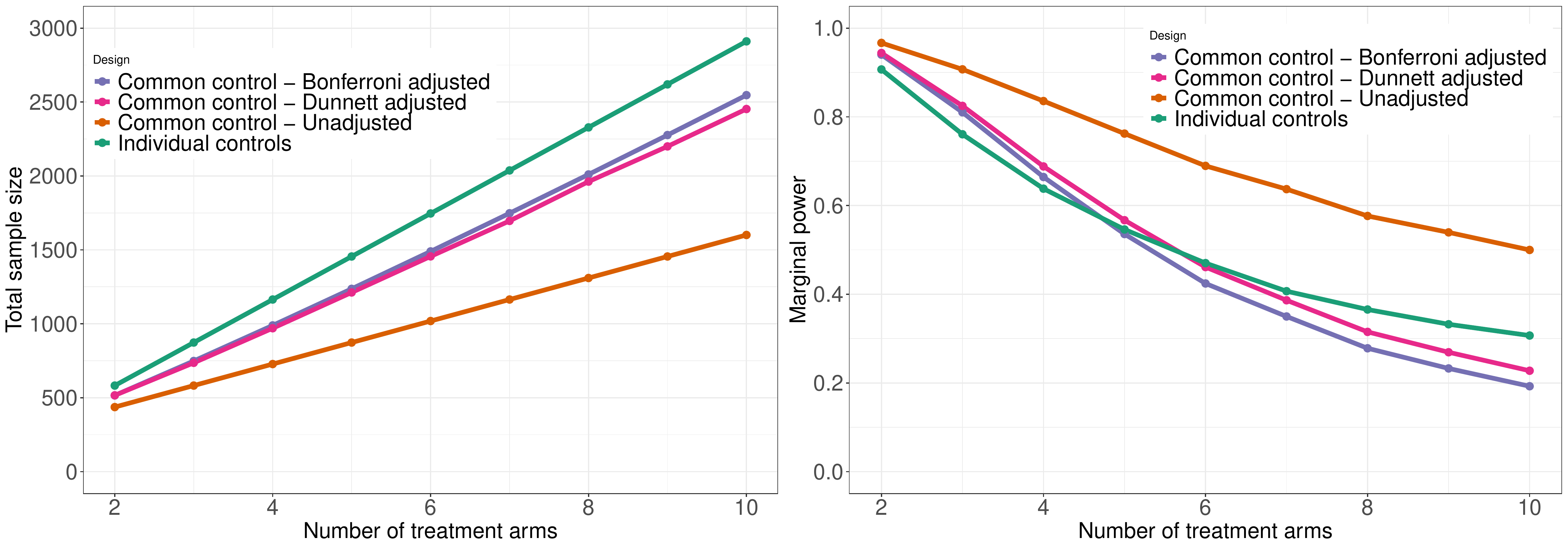}
  \caption{Left: Total sample size in platform trial with and without common control in order to obtain a marginal power of 90\% to detect a specific treatment effect with equal sample size per arm for $m=2,...,10$ treatment arms. Right: Marginal power with common control or with individual controls and fixed total sample size of $N_{CC} = N_{IC}= 600$ patients.}
  	\label{fig:MarginalPower_Fix}
\end{figure}

The overall sample size required for a marginal power of 90\% is always substantially lower with a common control compared to platform trials with individual controls (Figure~\ref{fig:MarginalPower_Fix}, left). As expected, the total sample size increases with increasing number of arms. Likewise, the marginal power with a fixed overall sample size is always larger with common controls in comparison to individual controls (Figure~\ref{fig:MarginalPower_Fix}, right) and decreases with an increasing number of arms. The higher marginal power of platform trials with common controls is a natural consequence of the fixed overall sample size. A common control leads to a larger sample size per comparison. For instance, with 3 treatment arms we allocate 150 patients to each arm in a platform trial with a common control, whereas with individual controls each arm contains 100 patients. 

If adjustment for multiplicity over the platform trial with common control is foreseen, a penalty in terms of power or sample size is to be paid. The total sample size in the platform trial with a common control increases. Nevertheless, the total sample size in order to obtain a marginal power of say 90\% to detect a specific treatment effect is higher with individual controls than with a common control. Results for an increasing number of arms are presented in Figure \ref{fig:MarginalPower_Fix} (left). As expected, Dunnett adjustment is more powerful than Boferroni adjustment, i.e. requires smaller sample sizes. However, the gain is relatively small.

In situations with fixed overall sample size and an increasing number of arms, or equivalently in situations were the sample size per arm gets very small, the multiplicity adjustments can become too strict. It thus might result in a lower power as compared to trials with individual controls (Figure \ref{fig:MarginalPower_Fix} (right)). With $5$ or more treatment arms the platform trial with a common control and a Bonferroni adjustment yields smaller (or equal) power than with individual controls, whereas with $6$ or more treatment arms Dunnett adjustments yields smaller (or equal) power than with individual controls. 
\cite{CharlesW.Dunnett.1955} showed that changing the equal allocation from 1:1:...:1 to 1:1:...:$\sqrt{m}$, allocating $\sqrt{m} \cdot n$ subjects to the common control group and $n$ subjects to the treatment arms in a trial with a common control results in an optimal power. By applying this optimal allocation, the number of patients in the control group and thus the total number of subjects per comparison increases. Furthermore, the correlation between test statistics decreases and therefore the cut point shifts to the right to 8 or more treatment arms (results not shown here).
This shows that in some circumstances, depending on the number of treatment arms and the total sample size, a platform trial with a common control can lose the advantages that result from using a common control over running a trial with individual controls per treatment arm, at least when multiplicity control is required.

\subsubsection*{All effective case: Comparison of disjunctive and conjunctive power}
\begin{figure}[ht]
	\centering
  \includegraphics[scale=0.35]{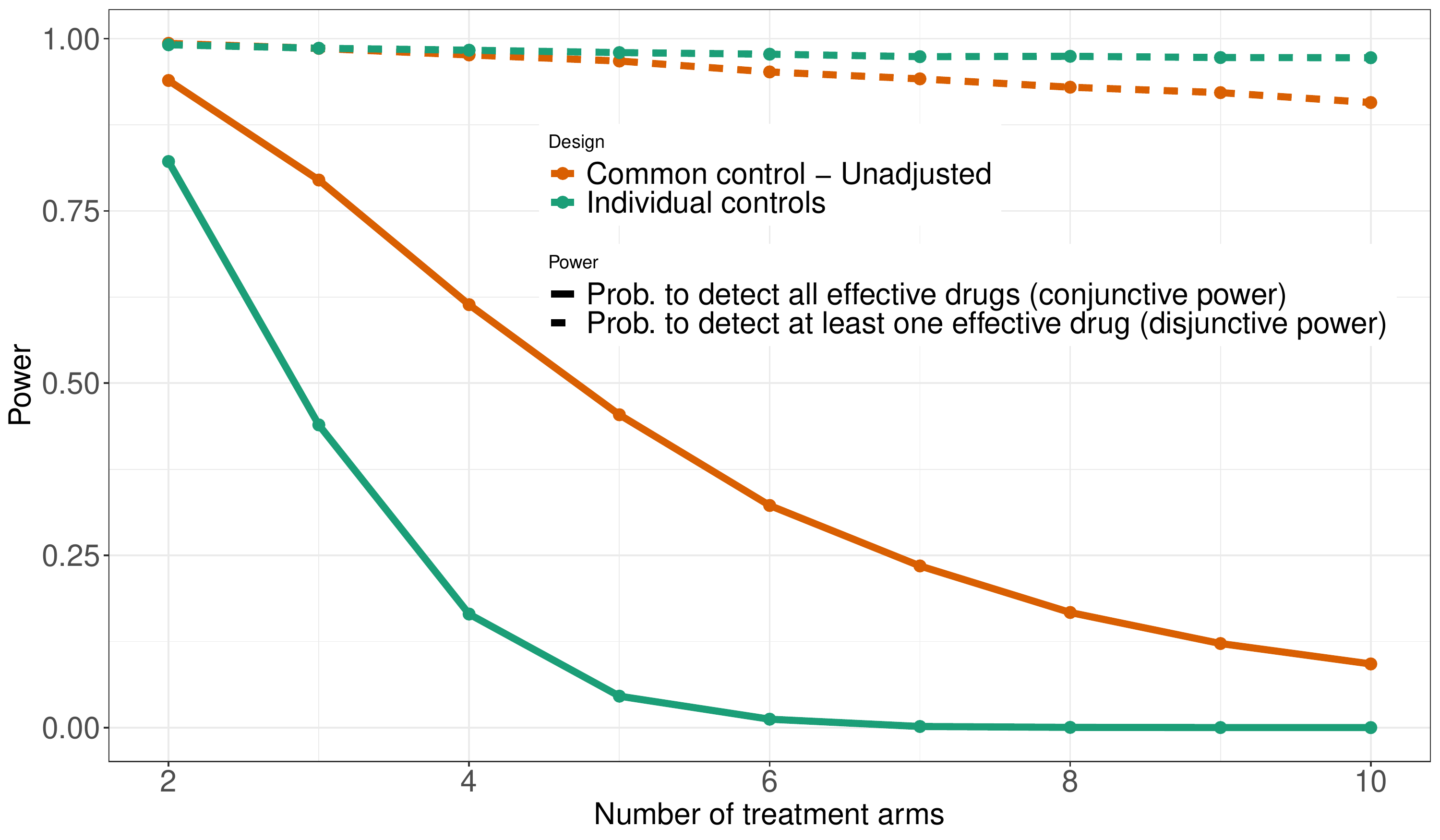}
  \caption{Disjunctive (probability to detect at least one effective drug) and conjunctive power (probability to detect all effective drugs) power for fixed sample size $N_{CC} = N_{IC} = 600$ patients.}
  	\label{fig:Disjunct_Conjunct}
\end{figure}

As with the type one error, in multiple testing situations, it might also be of interest to obtain an "overall" power over the platform, namely the disjunctive and conjunctive power. Figure \ref{fig:Disjunct_Conjunct} shows the probability to detect at least one effective drug (disjunctive power; dashed lines) as well as the probability to detect all effective drugs (conjunctive power; solid lines) for a platform trial with and without common control. While the probability to detect at least one effective drug is higher for trials with individual controls, the probability to detect all effective drugs is higher for a trial with a common control. Similarly to the explanation of the FWER and $k$-FWER, the use of a common control in a platform trial affects the two power concepts. If one effective drug is already detected, the likelihood to detect all other effective drugs is increased. On the other hand if a specific effective drug is not detected, it is more likely that also other effective drugs are not detected. The reason can be once more found in the common control, which can exhibit a random high or low. When Bonferroni or Dunnett adjustment is performed, both disjunctive and conjunctive power are decreased in comparison to the design with a common control without adjustment (results shown in online supplement). While the conjunctive power in the design with adjustment and common control is still greater than obtained with individual controls, the disjunctive power is lower since the platform trial without adjustment already results in smaller power than with individual controls.     

\subsection*{Flexible platform trial}
\subsubsection*{Null case: Comparison of error rates}
\begin{figure}[ht]
	\centering
  \includegraphics[scale=0.24]{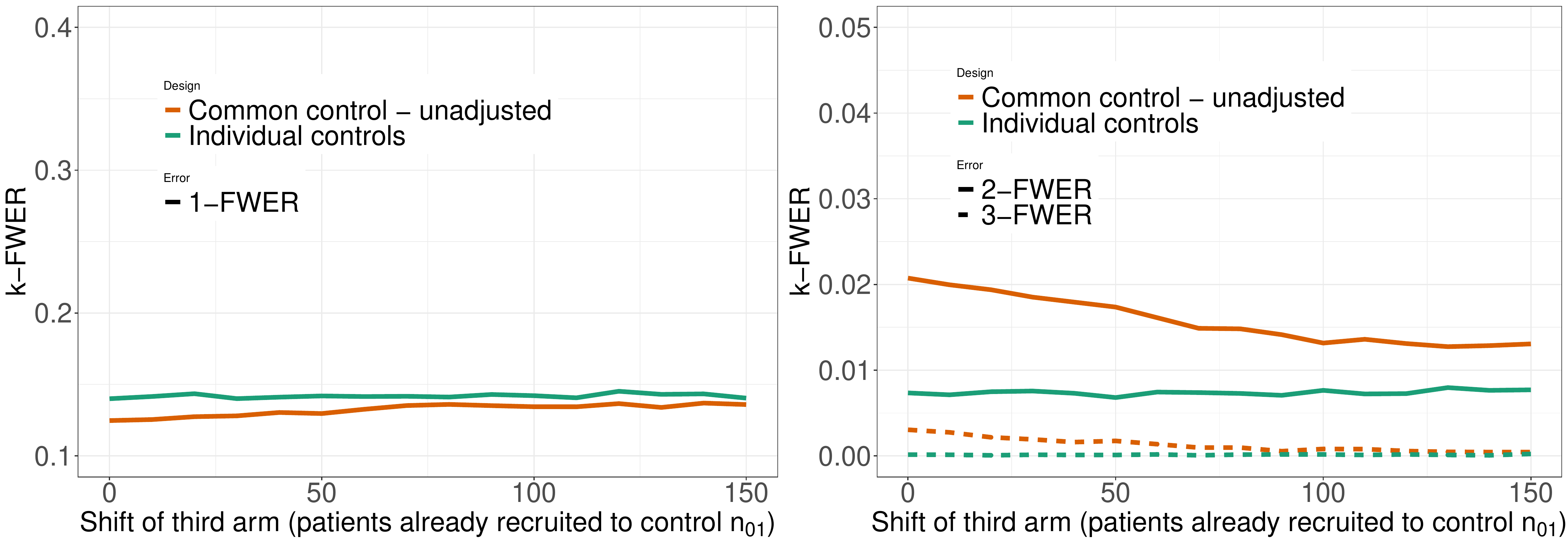}
  \caption{1-FWER (left), 2-FWER and 3-FWER (right) in a flexible platform trial of 3 treatments and a common control or individual controls. The x-axis represents the number of patients on the trial in the common control (of size $n_{01}$) when the third treatment enters the platform.}
  	\label{fig:FWER_Partial}
\end{figure}

In Figure \ref{fig:FWER_Partial}, the FWER and 2-FWER are shown with increasing number of patients on the platform when the third treatment arm joins the platform trial with a common control (of size $n_{01}$, see Figure~\ref{fig:Overlap} for definition of sample sizes). The later treatment 3 joins the platform, the smaller the number of common control patients between this and the other treatments. This results in a lower correlation and therefore the FWER in the platform trial with a common control slowly converges to the FWER as obtained with individual controls. The same is observed with the 2-FWER and the 3-FWER. The situation on the far left (0 patients recruited to the first two treatment arms and the common control group when treatment 3 enters the platform) represents a fixed platform, while the far right situation (150 patients recruited per arm when treatment 3 enters the platform) denotes no overlap of control group patients for the third treatment with the other two treatment groups - similarly to an independent setting. 
It is noted that the effect in the depicted scenario is not very strong, which is due to the relatively few treatment arms in the platform and hence the relatively small FWER increase with individual controls compared to common controls. The reason why the FWER, 2-FWER and 3-FWER do not fully overlap in this case is that there still exists a correlation between the first two treatments and hence a reduction in the FWER and an increase for the k-FWER for the platform trial with common control in total (result for 3-FWER not shown here). When Bonferroni or Dunnett adjustment is performed, the FWER is controlled and therefore also the $k$-FWER is below 0.05.

\subsubsection*{Single effective case: Comparison of marginal power}
\begin{figure}[ht]
	\centering
  \includegraphics[scale=0.23]{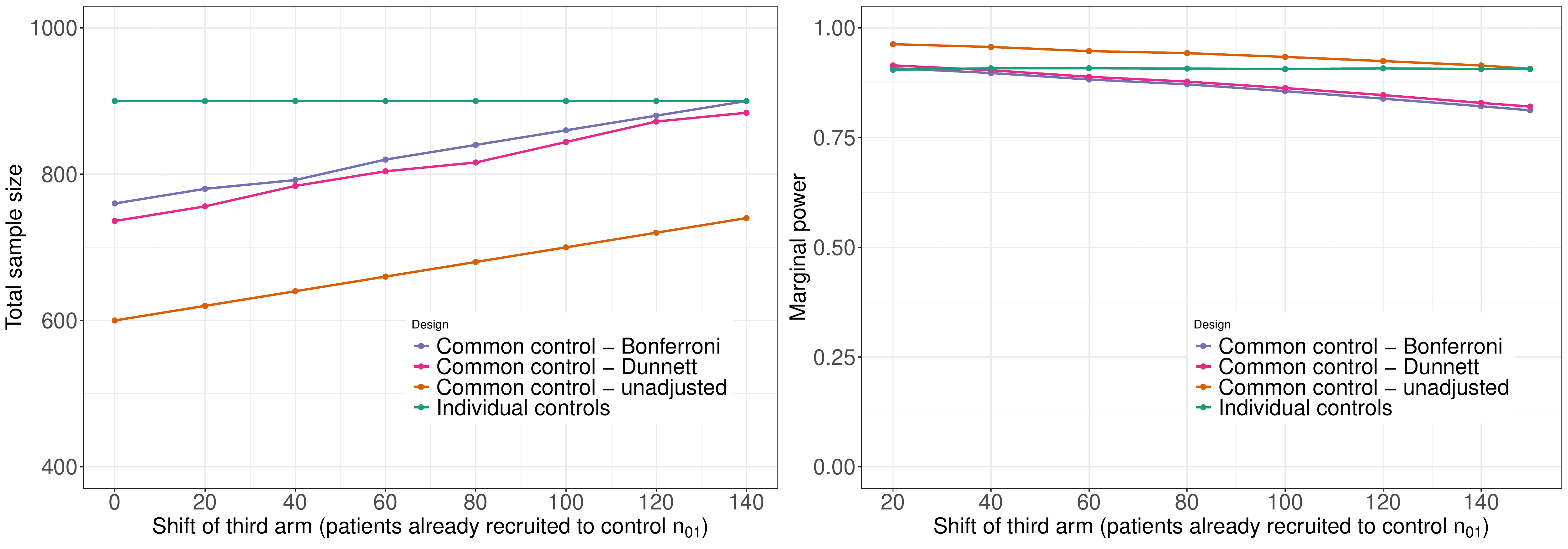}
  \caption{Left: Total sample size in platform trial of 3 treatments with and without common control in order to obtain a marginal power of 90\%. Right: Marginal power with fix comparison sample size of 300 patients. With increasing shift of treatment 3 to the platform with a common control, the use of the common control reduces.}
  	\label{fig:MarginalPower_FlexN}
\end{figure}

 In Figure \ref{fig:MarginalPower_FlexN} (left) the total sample size is shown with increasing number of patients already on the platform when the third treatment arm joins the platform with a common control in comparison to a platform trial with individual controls. The total sample size required to obtain a marginal power of 90\% remains the same for the platform trial with individual controls. In a platform trial with common controls the required total sample size increases with increasing number of patients already on the platform. Since only concurrently enrolled control patients are used for the comparison, the sample size in the control group and therefore the total sample size in the platform trial with a common control increases automatically by $n_{01}$, i.e. the non-concurrent control patients. When multiplicity adjustment is included, the increase is even higher and approaches the sample size obtained with individual controls and increasing $n_{01}$. 

Figure \ref{fig:MarginalPower_FlexN} (right) shows the marginal power with increasing shift of recruitment start (of size $n_{01}$) in treatment 3 when two treatment arms are already on the platform trial with a common control. As mentioned in Table \ref{tab:Shift_Recruitment}, it is assumed that the sponsor for treatment 3 can recruit a total of 300 patients and joins the platform at a later point in time. As long as the sponsor is able to use any common control patients, a sample size of $>150$ patients in the new active arm can be obtained with a common control. For the platform trial with individual controls, the sample size is equally split into 150 vs. 150 for the treatment and control arm.  
In the platform trial with a common control, with increasing shift of the third treatment arm, the number of common control patients reduces. With increasing shift, less patients can be used from the common control and the comparison-wise sample size converges to the one with individual controls. In the extreme case when treatment 3 joins the platform after 150 control patients were already recruited by the other treatment arms, no common control patients can be used. In this situation the same comparison-wise sample size is obtained as with individual controls, i.e. 150 vs. 150 patients. 

For the platform trial with a common control and multiplicity adjustment, a similar observation as in the fixed platform is obtained. The trial with a common control and adjustment results in a higher marginal power as compared to a trial with individual controls when a small shift in recruitment ($n_01$) is in place. However, with increasing shift, the number of common controls reduces, resulting in smaller comparison-wise sample sizes and therefore a reduction in power, while the adjustment further reduces the power. When more than 40 patients per treatment arm are already recruited in the platform trial with a common control, the marginal power is smaller with multiplicity adjustment than in the platform trial with individual controls. In this case, it is easy to show that one obtains a sample size of 187 for the treatment arm 3 as well as the concurrent control arm. Even though this sample size is higher than with an individual control (150 vs. 150), the adjustment overrules this benefit. For the sponsor, it would be therefore disadvantageous to join the ongoing platform trial with a common control if multiplicity adjustment is foreseen. 

\subsubsection*{All effective case: Comparison of disjunctive and conjunctive power}
\begin{figure}[ht]
	\centering
  \includegraphics[scale=0.35]{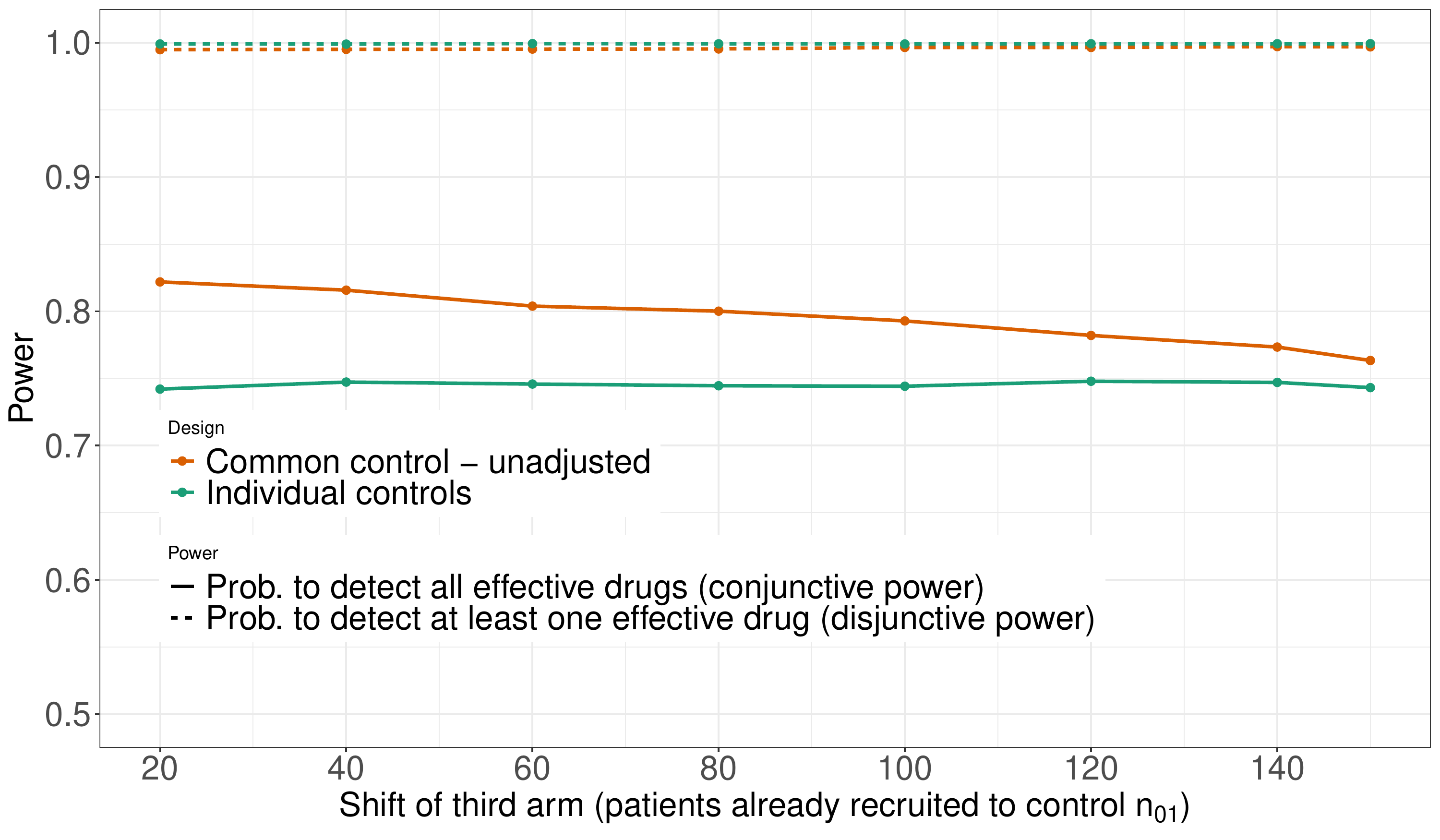}
  \caption{Disjunctive (probability to detect at least one effective drug) and conjunctive power (probability to detect all effective drugs) for fixed sample size}
  	\label{fig:Disjunct_Conjunct_Partial}
\end{figure}

In Figure \ref{fig:Disjunct_Conjunct_Partial}, the disjunctive (probability to detect at least one effective drug) and conjunctive power (probability to detect all effective drugs) are shown for a flexible platform trial with three treatment arms in an all effective case. On the x-axis the time of start of the third arm is displayed for the platform trial (of size $n_{01}$). With increasing shift of the arm, the number of common controls reduces and thus it can be seen that the conjunctive power of the trial with a common control converges to the power in the trial with individual controls, while there is only a minimal convergence of the disjunctive power. When Bonferroni or Dunnett adjustment is performed, both disjunctive and conjunctive power are decreased in comparison to the design with a common control without adjustment (results not shown here).    

\section*{Discussion}
\subsection*{Influence of multiplicity adjustment on sample size and power}
We have shown that, in many circumstances, platform trials with a common control can be beneficial in terms of sample size and power as compared to a trial with individual controls. Using a common control reduces the total required sample size for a given desired marginal power substantially. For a fixed total sample size, it positively affects the marginal power in a platform trial. 

It is clear that without multiplicity adjustment, the FWER of a platform trial is increased with multiple treatment arms, whether with or without a common control. In fact, the FWER inflation is even smaller in platform trials with a common control as compared to a trial with individual controls due to the correlation of test statistics. It converges to the FWER of platform trials with individual controls with decreasing correlation. The correlation can be decreased for example by allocating more patients to the control group, or by delaying the recruitment of one or more treatment arms. On the other hand, the opposite effect is observed for the k-FWER. Due to the use of a common control, the probability of two or more false positive findings is higher in a platform trial with a common control. However, the magnitude of the increase is much smaller. 

We further investigated how multiplicity adjustments in a platform trial with a common control affect the sample size or power. In many circumstances, multiplicity control in a design with a common control still leads to a gain in terms of the marginal power or sample size as compared to using individual controls. The advantages of a common control however, can be surpassed by the "penalty" of multiplicity adjustments in platform trials if the sample size used for each individual treatment-control comparison is small. Of course, it also depends on the number of treatment arms in the platform trial as this directly impacts the type 1 error adjustment. A high number of treatment arms and a small sample size per treatment-control comparison (even though it is higher than in individual trials) can lead to a smaller marginal power as obtained with individual controls with the same overall sample size. This is due to the strong "penalty" of the adjustment in combination with relatively small sample sizes. In this case, it might be worth running a trial with individual controls instead of joining the platform trial with a common control. Consequently, especially in situations with small overall achievable sample sizes, e.g. in rare diseases or pediatric investigations, in which streamlining clinical development might be of highest relevance, multiplicity adjustment in platform trials with a common control would would negate the advantage in terms of saving sample size. Therefore, the question remains, whether the mere use of a common control is reason enough to require multiplicity adjustment -- as platform trials with a common control can lose the advantages and benefits due to this penalty -- or if other considerations should guide the requirement for multiplicity control over the whole platform trial.

\subsection*{Current regulatory guidance}
With platform trials, the understanding of hypothesis family or hypotheses dependency in regulatory guidance documents might not be completely transferable from classical clinical trials to the more complex settings. Current guidelines on multiplicity issues such as the multiplicity guideline \citep{EMA.Multiplicity.2017} or the guideline on multiple endpoints \citep{FDA.Multiple.2017} require control of study-wise type I error rate for confirmatory claims - with emphasize on the study-wise control. Since these guidelines do not provide an exhaustive discussion of every issue and where apparently not written with complex platform trials in mind, it is not clear what might be expected for platform trials with and without common controls. Without further clarification, this could potentially lead to a overly conservative or even infeasible adjustment, either because the number of treatment arms is too high or because the number of arms for adjustments is not available at the beginning of the trial. Specific European guidance documents for complex clinical trials \citep{CTFG.2019,EMA.QnA.2022} mention the requirement for multiple treatment adjustments, but do not add further details on when this is necessary or how this could be achieved in practice. Similarly, the guideline on complex clinical trials from the FDA \citep{FDA.Complex.2020} requires the evaluation of "type I error probability control and power where applicable" without details when it would be applicable. FDA's guideline on master protocols does not discuss the type I error but refers to the guidance on adaptive designs, in which the type I error is discussed as the chance of erroneous conclusions in a \emph{trial} \citep{FDA.Master.2022}. In May 2021, FDA issued a guideline on master protocols of COVID-19 drug development and clarified that no multiplicity adjustments for multiple comparisons in platform or umbrella trials is needed. It is further acknowledged that potential multiple erroneous findings might arise \citep{FDA.COVID.2021}. It is not clear if platform trials for treatments other than COVID-19 drugs can follow the same approach. Even though not an official regulatory guidance, in \cite{Collignon.2020} personal views from the authors (of which several are or were members of the EMA Biostatistics Working Party) on statistical and regulatory considerations on confirmatory basket, umbrella and platform trials are shared. In the authors' view, the use of a common control does not necessarily require a multiplicity control, though corresponding regulator decisions are correlated. Further, the authors elaborate several situations in which adjustment might be needed or not. For example, the authors do not see a general need for an adjustment if the hypotheses are not related because these refer to different therapies. However, if several doses of the same treatment are evaluated, they considered that adjustment is needed.

\subsection*{Definition of a family of tests}
When planning a study, multiplicity adjustment is usually considered in order to control the family-wise error rate. This naturally leads to the determination of an adequate or reasonable family of tests to be controlled over. When defining a family of tests, the two extreme possibilities in a platform trial with a common control are (1) to define each treatment-control comparison as a separate family in which case no multiplicity adjustment is needed, or (2) as all treatment-control comparisons on the platform in which case one needs to control for multiplicity over all arms in the platform. \cite{Frane.2019} and \cite{Hung.2009} emphasized that the definition of a family is to some degree subjective but not completely arbitrarily. A family of tests in a platform trial can be defined based on different criteria, including but not limited to
\begin{itemize}
    \item the existence of dependence between the test statistics due to the use of a common control,
    \item the potential dependencies between the hypotheses (e.g. due to the use of the same or similar treatments),
    \item feasibility aspects (e.g. perpetual trials, special trial settings such as rare diseases or pediatric trials)
\end{itemize}
or an interaction of multiple aspects from above. While the first two aspects focus on dependencies/relatedness, the latter describes operational/feasibility criteria. Clearly these aspects are not solely of statistical nature, therefore the determination of a family of tests should involve cross-functional experts including but not limited to statisticians, clinicians as well as non-clinical colleagues.
Thus, the whole question of "when to adjust" can basically come down to what can be defined as a family. Thus, we have merely "postponed" the decision with this. In the following we will elaborate a bit more on the above listed criteria. The idea is not to provide concrete decision rules, but rather to highlight different aspects that can be considered alone or together when deciding whether and how to adjust.

\subsubsection*{Statistical dependency}
The use of a common control in a platform trial leads to dependencies between the test statistics. The impact on error rates and power has been discussed extensively in this publication. 
\cite{Rothman.1990} and \cite{Perneger.1998}, amongst others, argue that if we ran \emph{separate trials}, no multiplicity adjustment would be required. From an inferential point of view, running separate trials can be considered comparable to running a platform trial with \emph{individual controls}. In the latter case, one could easily use a two-step randomis7ation where subjects are first randomised to the sub-trial and then to the treatment or control arm of that trial rather than using a one-step randomisation. In both cases, however, the comparison within a subtrial would be in principle identical to that within a separate trial. The platform trial would only add an additional layer of randomisation to sub-trials (and hence in theory allowing randomised comparisons between sub-trials). It would not immediately impact on the type 1 error of the sub-trial itself, though, and multiplicity control over all sub-trials would hence not be needed. As we have seen, the FWER in platform trials with common controls is even lower than in platform trials with individual controls. From this perspective it seems counter-intuitive to request type 1 error control over all arms in a platform trial just because of a common control. 

\subsubsection*{Dependency between hypotheses}
If we consider different doses of the same drug, or quasi identical compounds (with only minimal differences but the same mechanism of action) in a platform trial, another form of dependency, namely the dependencies between the hypothesis or and hence between the decision whether one of the doses or one of the treatments is decided to be efficacious or not, is added to the discussion. This dependency needs to be discussed in the context of the trial based on biological, chemical and clinical considerations. Multiple doses of the same treatment lead to dependent hypotheses in the sense tat the event that one of the doses is effective is dependent on the event that another dose is effective. Therefore, testing two doses of a drug (in one trial) inflates the type 1 error for that drug in a meaningful way as we have multiple chances to (erroneously) conclude that the drug is effective. It is much more difficult to agree whether treatments with identical or similar mode of action also lead to dependent hypotheses, or whether treatments from the same sponsor should always be considered dependent as they increase the chance of the sponsor for an (erroneously) effective treatment. 

For the determination of independent hypotheses, a justification should be provided based on cross-functional discussions as mentioned above. For example biological independence could be justified with details on different modes of action or different biological targets. But solely stating that each treatment-control comparison leads to independent submissions or label claims is usually not enough. The independence of research questions needs to be thoroughly justified. If one treatment is declared superior in an ongoing trial, does this affect the decision of continuing or stopping another treatment? If it can be clearly shown that the hypotheses are independent of each other, adjustment is likely not needed.

\subsubsection*{Feasibility aspects}
Besides the discussion on the different independence aspects for the appropriate family definition, feasibility is one of the important aspects to be considered. Especially, in perpetual trials, it might be too conservative or impossible to adjust over all treatments that enter the trial at some point (prior and future arms), even years after other treatments have stopped recruitment already. I-SPY 2 is, for example, an ongoing platform trial of treatments for breast cancer with more than 20 treatments (both ongoing or already completed), spanning a time frame of more than 10 years. However, not all treatments were concurrently on the platform - only up to around 5 treatment arms were concurrently seen on the I-SPY 2 agent timeline \citep{ISPY2.2022}. Whether to adjust for 5, for 20 or for a pre-defined maximum number of treatments will have a huge impact on the multiplicity adjustment and therefore the decision to join a platform. Not only is the adjustment very conservative or infeasible but might be ethically questioned. One might argue that if no control data is common among the treatments, it might be reasonable to restrict the definition of a family to those treatments that are concurrently recruiting and thus share control group data. However, as seen above, not only the common control group alone determines the need for adjustment but also the question on other dependencies. If these non-concurrent treatments are of the same drug with different doses, adjustment is likely needed. 

Furthermore, in special cases such as a rare disease, in pediatric settings or in exploratory trials different requirements might be discussed based on further aspects such as feasibility and objective of the platform trial. Even if multiplicity control might be needed for a platform trial in a rare disease or pediatric population, the question on feasibility must be considered in addition. Generally, requirements for multiplicity control in exploratory trials are less stringent as the outcome from such trials are usually not used used for confirmatory claims. However, in some cases, exceptional results might still result in a submission leading to concerns with the interpretation and result if no adjustment was in place.   

Finally, one should be reminded that the use of a common control already leads to a reduction in the total sample size as compared to individual controls and in many circumstances even after multiplicity control a smaller total sample size is obtained. The additional "gain" if no multiplicity control is performed with a common control should be weighted against the risk of an increased $k$-FWER.

\section*{Summary and limitations}
For our simulations, we have only considered the simple  situation where the primary analysis was performed \emph{after all} treatment arms stopped recruitment, and the total number of treatment arms for adjustments was known. In our simulations, the time of joining the trial would have not affected the power if the effective arm would have been one of the arms already on the trial. For the arms already on the platform, the treatment-control sample size is not affected by the late joining of the third arm and multiplicity control was fixed given the number of arms for adjustment is known. Therefore, we only considered the effective treatment arm to join the flexible platform late and allowed for an increase in the pairwise sample size for this arm and the concurrent control for the analysis of power. In practice, these simplistic assumptions are usually not true. Especially the total number of treatment arms in case of perpetual platform trials might not be available at the start of the trial and one does not wait with the analysis until all arms were closed. Waiting with the analysis poses a commercial and ethical issue, on the other hand it would prevent unblinding or biasing the ongoing study. In practice, adjustment for multiplicity would change at the time a new treatment joins the trial. Several methods have been proposed in the literature for adjustment in such cases, such as the online control of error rates \citep{Tian.2021}. These methods allow for adjustments based on the previous hypotheses tests and do not require the knowledge of future hypotheses tests. Since each hypothesis test is dependent on the previous rejections, the time of joining the platform trial determines how much alpha is left for testing. Therefore, we have focused on the traditional adjustment methods in our simulations. The online control of error rates is thus out of scope. Though the FWER is smaller with a common control, the $k$-FWER ($k \geq 2$) is higher as compared to individual controls. In case no multiplicity control for the FWER is acceptable, the control of $k$-FWER might be considered as an alternative. This would ensure that the false multiple error rate obtained with a common control is not higher than obtained with individual controls. Methods for the control of the $k$-FWER were proposed by \cite{Lehmann.2005} and \cite{Keselman.2011}. Future research might evaluate the impact of the $k$-FWER control on the operational characteristics.

The aim of our simulations was to aid discussions on the need for multiplicity adjustments and its impact in platform trials with a common control and therefore provide high level results for discussions on the general need with regards to a common control. In addition to the potential operational advantage of a platform trial and the use of a common control, in many situations, a multiplicity adjustment can still lead to a lower total sample size as compared to using individual controls or running individual trials. Overall, not only consensus among researchers is currently missing but also further guidance from regulators is pending. As noted above, the use of a common control alone cannot determine the need for adjustment. Thus, we propose that a justification based on several concepts for the determination of an appropriate family should be provided with the overall goal to not erroneously conclude effective drugs and its consequences in a meaningful way. A case-to-case discussion for each new platform trial and treatment arm needs to be initiated and justification based on statistical, biological and chemical arguments need to be provided.

\section*{Appendix}
Proof of equation \eqref{eq.corr.1}:\\
Let
\begin{align*}
 Z_j^* := \frac{\bar{X^j} - \bar{X^{0j}}}{\sqrt{\frac{1}{n_{j}}+\frac{1}{n_0^j}}} \sim \mathcal{N}(0,1),
\end{align*}
be the test statistic of the comparison of treatment $j$ with the common control group in a flexible platform trial. Furthermore we have 
\begin{align*}
I_j &\coloneqq  \Big\{ i = 1,\ldots,n_0: \text{ Patient } i \text{ in control arm, concu.} \text{ randomized to arm } j \Big\}, \\
n_0^j &\coloneqq \#I_j, j = 1, \ldots, m \text{ and } n_0^{j,j'}:=\#\left(I_j \cup I_{j'}\right) \text{ for } j,j' = 1, \ldots, m. \\
\end{align*}
For the correlation $\rho_{j, {j^\prime}}$ between the test statistics $Z_j^*$ and $Z_{j'}^*$ for $j,j'=1,\ldots,m$ and $j\neq j'$ one receives by standard calculations
\begin{align*}
    \rho_{j, {j^\prime}} &= \text{ Corr}(Z_j^*, Z_{j^\prime}^*) = \frac{\text{Cov}(Z_j^*, Z^*_{j^\prime})}{\sqrt{\text{Var}(Z_j^*) \text{ Var}(Z^*_{j^\prime})}} = \text{Cov}(Z_j^*, Z^*_{j^\prime}) \\
    & = \frac{1}{\sqrt{\frac{1}{n_j} + \frac{1}{n_{0}^j}}}  \frac{1}{\sqrt{\frac{1}{n_{j^\prime}} + \frac{1}{n_{0}^{j^\prime}}}} \cdot \text{Cov}(\bar{X_j} - \bar{X}^{0j}, \bar{X}_{j^\prime} - \bar{X}^{0j^\prime}).
\end{align*}
Since the second factor can be rewritten to
\begin{align*}
    &\text{Cov}(\bar{X_j} - \bar{X}^{0j}, \bar{X}_{j^\prime} - \bar{X}^{0j^\prime}) \\
    & = \frac{1}{2} \Big[\text{Var}(\bar{X_j} - \bar{X}^{0j} -  \bar{X}_{j^\prime} - \bar{X}^{0j^\prime}) - \text{Var}(\bar{X_j} - \bar{X}^{0j}) - \text{Var}(\bar{X}_{j^\prime} - \bar{X}^{0j^\prime})\Big] \\ 
    & = \frac{1}{2} \Big[ \text{Var}(\bar{X}^{0j} + \bar{X}^{0j^\prime}) - \text{Var}(\bar{X}^{0j}) - \text{Var}(\bar{X}^{0j^\prime}) \Big] \\ 
    & = \frac{1}{2} \Bigg[ \text{Var} \bigg(\frac{1}{\# I_j} \sum_{i \in I_j} X_i^0 + \frac{1}{\# I_{j^\prime}} \sum_{i \in I_{j^\prime}} X_i^0 \bigg) - \frac{1}{n_0^j} - \frac{1}{n_0^{j^\prime}} \Bigg] \\
    & = \begin{aligned}[t]\frac{1}{2} \Bigg[ \text{Var} \bigg(&\frac{1}{n_0^j} \Big( \sum_{i \in I_j \setminus \{I_j \cap I_{j^\prime}\}} X_i^0 + \sum_{i \in \{ I_j \cap I_{j^\prime} \}} X_i^0 \Big) \\ 
    + & \frac{1}{n_0^{j^\prime}} \Big( \sum_{i \in I_{j^\prime} \setminus \{I_j \cap I_{j^\prime}\}} X_i^0 + \sum_{i \in \{ I_j \cap I_{j^\prime} \}} X_i^0 \Big) \bigg) - \frac{1}{n_0^j} - \frac{1}{n_0^{j^\prime}} \Bigg]\end{aligned}\\
    & = \frac{1}{2} \Bigg[ \frac{n_0^{j,j'} - n_0^{j^\prime}}{(n_0^j)^2} + \frac{n_0^{j,j'} - n_0^j}{(n_0^{j^\prime})^2} + \bigg( \frac{1}{n_0^j} + \frac{1}{n_0^{j^\prime}} \bigg)^2 (n_0^j + n_0^{j^\prime} - n_0^{j,j'}) - \frac{1}{n_0^j} - \frac{1}{n_0^{j^\prime}} \Bigg] \\
    & = \begin{aligned}[t] \frac{1}{2} \Bigg[ &\frac{n_0^{j,j'} - n_0^{j^\prime}}{(n_0^j)^2} + \frac{n_0^{j,j'} - n_0^j}{(n_0^{j^\prime})^2} + \bigg( \frac{1}{(n_0^j)^2} + \frac{2}{n_0^j n_0^{j^\prime}} + \frac{1}{(n_0^{j^\prime})^2}  \bigg) \cdot ( n_0^j + n_0^{j^\prime} - n_0^{j,j'} ) \\ 
    &-\frac{1}{n_0^j} - \frac{1}{n_0^{j^\prime}} \Bigg] \end{aligned} \\
    & = \begin{aligned}[t] \frac{1}{2} \Bigg[ &\frac{n_0^{j,j'} - n_0^{j^\prime}}{(n_0^j)^2} + \frac{n_0^{j,j'} - n_0^j}{(n_0^{j^\prime})^2} + \frac{n_0^j + n_0^{j^\prime} - n_0^{j,j'}}{(n_0^{j})^2}  +\frac{n_0^j + n_0^{j^\prime} - n_0^{j,j'}}{(n_0^{j^\prime})^2} + \frac{2 (n_0^j + n_0^{j^\prime} - n_0^{j,j'} ) }{n_0^j n_0^{j^\prime}} \\ 
    & - \frac{1}{n_0^j} - \frac{1}{n_0^{j^\prime}} \Bigg] \end{aligned} \\
    & = \frac{n_0^j + n_0^{j^\prime} - n_0^{j,j'}}{n_0^j n_0^{j^\prime}},
\end{align*}
we receive altogether
\begin{align*}
    \rho_{j, {j^\prime}} &= \frac{1}{\sqrt{\frac{1}{n_j} + \frac{1}{n_{0}^j}}\sqrt{\frac{1}{n_{j^\prime}} + \frac{1}{n_{0}^{j^\prime}}}} \frac{n_0^j + n_0^{j^\prime} - n_0^{j,j'}}{n_0^j n_0^{j^\prime}}.
\end{align*}


\begin{backmatter}

\section*{Acknowledgements}
The present work was performed in (partial) fulfillment of the requirements for obtaining the degree Dr. rer. biol. hum.
The views expressed in this article are the personal views of the author(s) and may not be understood or quoted as being made on behalf of or reflecting the position of the regulatory agencies or organisations with which the authors are affiliated.
The authors also wish to thank Hue Kästel for the help in preparing the appendix.

\section*{Funding}
EU-PEARL (EU Patient-cEntric clinicAl tRial pLatforms) project has received funding from the Innovative Medicines Initiative (IMI) 2 Joint Undertaking (JU) under grant agreement No 853966. This Joint Undertaking receives support from the European Union’s Horizon 2020 research and innovation programme and EFPIA and Children’s Tumor Foundation, Global Alliance for TB Drug Development non-profit organisation, Springworks Therapeutics Inc. This publication reflects the authors’ views. Neither IMI nor the European Union, EFPIA, or any Associated Partners are responsible for any use that may be made of the information contained herein.




\section*{Competing interests}
The authors declare that they have no competing interests.





\typeout{} 

\bibliographystyle{bmc-mathphys} 
\bibliography{bmc_article}      
\nocite{label}










\end{backmatter}
\end{document}